\documentclass[11pt,letterpaper]{article}

\newcommand{\captionfonts}{\small}

\makeatletter  
\long\def\@makecaption#1#2{%
  \vskip\abovecaptionskip
  \sbox\@tempboxa{{\captionfonts #1: #2}}%
  \ifdim \wd\@tempboxa >\hsize
    {\captionfonts #1: #2\par}
  \else
    \hbox to\hsize{\hfil\box\@tempboxa\hfil}%
  \fi
  \vskip\belowcaptionskip}
\makeatother   

\usepackage[utf8]{inputenc}
\usepackage[usenames,dvipsnames]{color}
\usepackage{enumerate}
\usepackage{amsfonts, amsmath, amssymb, mathrsfs}
\usepackage[amsmath,hyperref,amsthm]{ntheorem}
\usepackage{tikz}
\usetikzlibrary{shapes,snakes,matrix}

\def\beginskinny{\begin{center}\begin{minipage}[l]{0.85\textwidth}}
\def\endskinny{\end{minipage}\end{center}}

\def\calG{\mathcal{G}}
\def\calP{\mathcal{P}}
\def\calC{\mathcal{C}}
\def\calS{\mathcal{S}}
\def\calM{\mathcal{M}}
\def\calN{\mathcal{N}}

\begin{document}
\title{$k$-means requires exponentially many iterations even in the plane}
\author{Andrea Vattani\\{\small University of California, San Diego} \\{\small \tt avattani@ucsd.edu}}
\date{}
\maketitle


\thispagestyle{empty}

\abstract{
The $k$-means algorithm is a well-known method for partitioning $n$
points that lie in the $d$-dimensional space into $k$ clusters. Its main
features are simplicity and speed in practice. Theoretically, however,
the best known upper bound on its running time (i.e. $O(n^{kd})$)
can be exponential in the number of points. Recently, Arthur and
Vassilvitskii \cite{arthur} showed a super-polynomial worst-case analysis,
improving the best known lower bound from $\Omega(n)$ to
$2^{\Omega(\sqrt{n})}$ with a construction in $d=\Omega(\sqrt{n})$
dimensions.
In \cite{arthur} they also conjectured the existence of
super-polynomial lower bounds for any $d\ge 2$.

Our contribution is twofold: we prove this conjecture and we improve
the lower bound, by presenting a simple construction in the
plane that leads to the exponential lower bound $2^{\Omega(n)}$. 
}

\newpage 

\setcounter{page}{1}

\section{Introduction}
The $k$-means method is one of the most widely used algorithms for
geometric clustering. It was originally proposed by Forgy in 1965 \cite{forgy} 
and McQueen in 1967 \cite{mcqueen}, and is often known as Lloyd's algorithm \cite{lloyd}. 
It is a local search algorithm and partitions $n$ data points
into $k$ clusters in this way: seeded with $k$ 
initial cluster centers, it assigns every data point to its closest
center, and then recomputes the new centers as the means (or centers of mass) of
their assigned points. This process of assigning data points and
readjusting centers is repeated until it stabilizes.

Despite its age, $k$-means is still very popular today and is considered
``by far the most popular clustering algorithm used in scientific and
industrial applications'', as Berkhin remarks in his survey on
data mining \cite{berkhin}. 
Its widespread usage extends over a variety of different areas, such as
artificial intelligence, computational biology, computer graphics,
just to name a few (see \cite{agarwal, gibou}). It is particularly
popular because of its simplicity and observed speed: as Duda et
al. say in their text on pattern classification \cite{duda}, ``In
practice the number of iterations is much less than the number of
samples''.

\smallskip

Even if, in practice, speed is recognized as one of $k$-means' main
qualities (see \cite{kanungo} for empirical studies), on
the other hand there are a few theoretical bounds on its
worst-case running time and they do not corroborate this feature. 

An upper bound of $O(k^n)$ can be trivially established since it can
be shown that no clustering occurs twice during the course of the
algorithm. In \cite{inaba}, Inaba et al. improved this bound to
$O(n^{kd})$ by counting the number of Voronoi partitions of $n$ points in
$\mathbb{R}^d$ into $k$ classes. Other bounds are known for some special
cases. Namely, Dasgupta \cite{dasgupta} analyzed the case $d=1$, proving an upper bound of
$O(n)$ when $k<5$, and a worst-case lower bound of $\Omega(n)$. Later, Har-Peled and Sadri
\cite{harpeled}, again for the one-dimensional case, showed an upper
bound of $O(n\Delta^2)$ where $\Delta$ is the spread of the point set
(i.e. the ratio between the largest and the smallest pairwise
distance), and conjectured that $k$-means
might run in time polynomial in $n$ and $\Delta$ for any $d$.  

The upper bound $O(n^{kd})$ for the general case has not been improved
since more than a decade, and this suggests that it might be not far
from the truth. Arthur and Vassilvitskii \cite{arthur} showed that
$k$-means can run for super-polynomially many iterations, improving
the best known lower bound from $\Omega(n)$ \cite{dasgupta} to
$2^{\Omega(\sqrt{n})}$. 
Their contruction lies in a space with $d=\Theta(\log n)$ dimensions,
and they leave an open question about the performance of $k$-means for a
smaller number of dimensions $d$, conjecturing the existence of
superpolynomial lower bounds when $d>1$. Also they show that their
construction can be modified to have low spread, disproving the
aforementioned conjecture in \cite{harpeled} for $d=\Omega(\log n)$.

A more recent line of work that aims to close the gap between
practical and theoretical performance makes use of the smoothed
analysis introduced by Spielman and Teng \cite{spielman}. Arthur and
Vassilvitskii \cite{arthur2} proved a smoothed upper bound of
poly$(n^{O(k)})$, recently improved to poly($n^{O(\sqrt{k})}$) by Manthey and
R\"oglin \cite{manthey}. 

\subsection{Our result}
In this work we are interested in the performance of $k$-means
in a low dimensional space. We said it is conjectured \cite{arthur} that there exist instances
in $d$ dimensions for any $d\ge 2$, for which $k$-means runs for a super-polynomial
number of iterations.

Our main result is a construction in the plane ($d=2$) for which
$k$-means requires exponentially many iterations to stabilize. Specifically, we
present a set of $n$ data points lying in $\mathbb{R}^2$, and a set of
$k=\Theta(n)$ adversarially chosen cluster centers in $\mathbb{R}^2$, for which the algorithm runs for
$2^{\Omega(n)}$ iterations. This proves the aforementioned conjecture
and, at the same time, it also improves the best known lower bound from
$2^{\Omega(\sqrt{n})}$ to $2^{\Omega(n)}$. 
Notice that the exponent is optimal disregarding logarithmic factor, since
the bound for the general case $O(n^{kd})$ can be rewritten as $2^{O(n\log n)}$
when $d=2$ and $k=\Theta(n)$.
For any $k=o(n)$, our lower bound easily translates to
$2^{\Omega(k)}$, which, analogously, is almost optimal since the
upper bound is $2^{O(k\log n)}$. 

A common practice for seeding $k$-means is to choose the initial centers 
as a subset of the data points. We show that even in this
case (i.e. cluster centers adversarially chosen among the data points),
the running time of $k$-means is still exponential.

Also, using a result in \cite{arthur}, our construction can be modified to
an instance in $d=3$ dimensions having low spread for which $k$-means
requires $2^{\Omega(n)}$ iterations, which disproves the
conjecture of Har-Peled and Sadri \cite{harpeled} for any $d\ge 3$. 

Finally, we observe that our result implies that the smoothed analysis
helps even for a small number of dimensions, since the best smoothed
upper bound is $n^{O(\sqrt{k})}$, while our lower bound is
$2^{\Omega(k)}$ which is larger for $k=\omega(\log^2 n)$. In other
words, perturbing each data point and then running $k$-means would
improve the performance of the algorithm.

\section{The $k$-means algorithm}
The $k$-means algorithm allows to partition a set $X$ of $n$ points in
$\mathbb{R}^d$ into $k$ clusters. It is seeded with any initial set of
$k$ cluster centers in $\mathbb{R}^d$, and given
the cluster centers, every data point is assigned to the
cluster whose center is closer to it. The name ``$k$-means'' refers to
the fact that the new position of a center is computed as the center
of mass (or mean point) of the points assigned to it.

\smallskip

A formal definition of the algorithm is the following:
\begin{enumerate}
\item[0.] Arbitrarily choose $k$ initial centers $c_1,c_2,\ldots, c_k$.
\item[1.] For each $1\le i\le k$, set the cluster $C_i$ be the set of points in
  $X$ that are closer to $c_i$ than to any $c_j$ with $j\neq i$.
\item[2.] For each $1\le i\le k$, set $c_i = \frac{1}{|C_i|}\sum_{x\in
C_i} x$, i.e the center of mass of the points in $C_i$.
\item[3.] Repeat steps 1 and 2 until the clusters $C_i$ and the
  centers $c_i$ do not change anymore. The partition of $X$ is the set
  of clusters $C_1, C_2, \ldots, C_k$.
\end{enumerate}

Note that the algorithm might incur in two possibile ``degenerate''
situations: the first one is when no points are assigned to a
center, and in this case that center is removed and we will
obtain a partition with less than $k$ clusters. The other
degeneracy is when a point is equally close to more than one center,
and in this case the tie is broken arbitrarily.

We stress that when $k$-means runs on our constructions, it does not
fall into any of these situations, so the lower bound does not exploit
these degeneracies. 
 
\smallskip

Our construction use points that have constant integer weights. 
This means that the data set that $k$-means will take in input is actually a
multiset, and the center of mass of a cluster $C_i$ (step 2 of $k$-means) is
computed as $\sum_{x\in C_i} w_x x /\sum_{x\in C_i}w_x$, where $w_x$ is the weight of $x$. 
This is not a restriction since integer weights in the range $[1,C]$ 
can be simulated by blowing up the size of the data set by at most $C$: 
it is enough to replace each point $x$ of weight $w$ with a set of $w$ distinct points (of unitary weight) 
whose center of mass is $x$, and so close each other that the behavior of $k$-means 
(as well as its number of iterations) is not affected.

\section{Lower bound}
In this section we present a construction in the plane for which
$k$-means requires $2^{\Omega(n)}$ iterations. We start with some high
level intuition of the construction, then we give some definitions
explaining the idea behind the construction, and finally we proceed to the
formal proof. 

In the end of the section, we show a couple of extensions:
the first one is a modification of our construction so that the initial set of 
centers is a subset of the data points, and the second one describes
how to obtain low spread.     

A simple implementation in Python of the lower bound is available at
the web address \texttt{http://www.cse.ucsd.edu/\~{}avattani/k-means/lowerbound.py}

\subsection{High level intuition}
The idea behind our construction is simple and can be related to the saying
``Who watches the watchmen?'' (or the original latin phrase ``Quis
custodiet ipsos custodes?''). 

Consider a sequence of $t$ watchmen $W_0,W_1, \ldots, W_{t-1}$. 
A ``day'' of a watchman $W_i$ ($i>0$) can be described as
follows (see Fig.~1): $W_i$ watches $W_{i-1}$, waking it up once it falls asleep,
and does so twice; afterwards, $W_i$ falls asleep itself. 
The watchman $W_0$ instead will simply fall asleep directly after it has been
woken up. Now if each watchman is awake in the beginning of this
process (or even just $W_{t-1}$), it is
clear that $W_0$ will be woken up $2^{\Omega(t)}$ times by the time that
every watchman is asleep.

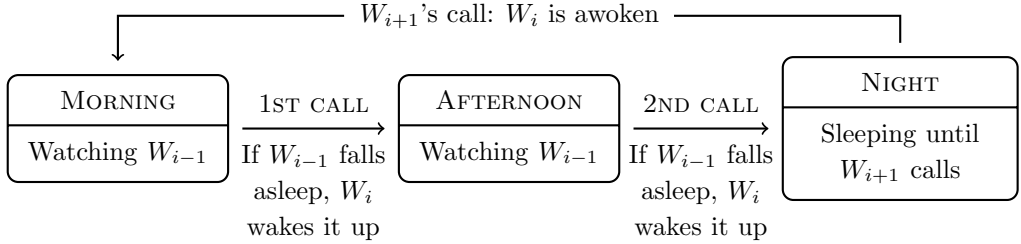
\begin{figure}[t]
\begin{center}
\begin{tikzpicture}[
    auto, node distance=5.2cm,
    myarrow/.style={thick,draw=black, shorten >=5pt, shorten <=5pt},
    myarrow2/.style={thick,draw=black},
    kant/.style={text width=2cm, text centered, sloped},
    every node/.style={text ragged, inner sep=2mm},
    punkt/.style={rectangle, rounded corners,draw=black, thick }
    ]
\newcommand{\scalefont}[1]{{\small #1}}
\node[name=morning,punkt,text width=6.5em, rectangle split, rectangle split parts=2, text centered] {
 \scalefont{\textsc{Morning}}
 \nodepart{second}
 \scalefont{Watching $W_{i-1}$}
};
\node [name=afternoon, right of=morning, punkt, text width=6.5em, rectangle split, rectangle split parts=2, text centered] {
  \scalefont{\textsc{Afternoon}}
  \nodepart{second}
  \scalefont{Watching $W_{i-1}$}
};
\node [name=night, right of=afternoon, punkt, text width=7em, rectangle split, rectangle split parts=2, text centered] {
  \scalefont{\textsc{Night}}
  \nodepart{second}
  \scalefont{Sleeping until $W_{i+1}$ calls}
};

\node[name=phony, above of=afternoon, node distance=1.5cm]{\scalefont{$W_{i+1}$'s call: $W_i$ is awoken}};

\draw [->, myarrow] (morning) -- node[kant, below]  {\scalefont{If $W_{i-1}$ falls asleep, $W_i$ wakes it up}} node[kant, above] {\scalefont{\textsc{1st call}}}(afternoon);
\draw [->, myarrow] (afternoon) -- node[kant, below]  {\scalefont{If $W_{i-1}$ falls asleep, $W_i$ wakes it up}} node[kant, above] {\scalefont{\textsc{2nd call}}}(night);
\draw[myarrow2, shorten <=5pt] (night) |- (phony);
\draw[->,myarrow2, shorten >=5pt] (phony) -| (morning);
     
\end{tikzpicture}

\parbox{5.5in}{\caption{\small The ``day'' of the watchman $W_i$, $i>0$.}}
\end{center}
\end{figure}

\smallskip

In the construction we have a sequence of gadgets $\calG_0, \calG_1, \ldots
\calG_{t-1}$, where all gadgets $\calG_i$ with $i>0$ are identical except for the
scale. Any gadget $\calG_i$ ($i>0$) has a fixed number of points and two centers, and
different clusterings of its points will model which stage of the day
$\calG_i$ is in. The clustering indicating that $\calG_i$ ``fell asleep'' has
one center in a particular position $S^*_i$. 

In the situation when $\calG_{i+1}$ is awake and $\calG_{i}$ falls
asleep, some points of $\calG_{i+1}$ will be assigned temporarily to the
$\calG_{i}$'s center located in $S^*_{i}$; in the next step this center will
move so that in one more step the initial clustering (or ``morning clustering'') of $\calG_{i}$ is
restored: this models the fact that $\calG_{i+1}$ wakes up $\calG_{i}$.

Note that since each gadget has a constant number of centers, we can
build an instance with $k$ clusters that has $t=\Theta(k)$ gadgets, for
which $k$-means will require $2^{\Omega(k)}$ iterations. 
Also since each gadget has a constant number of points, we can build
an instance of $n$ points and $k=\Theta(n)$ clusters with
$t=\Theta(n)$ gadgets. This will imply a lower bound of
$2^{\Omega(n)}$ on the running time of $k$-means.

\subsection{Definitions and further intuition}\label{defs}
For any $i>0$, the gadget $\calG_i$ is a tuple $(\calP_i,\calC_i, r_i, R_i)$ where
$\calP_i\subset \mathbb{R}^2$ is the set of points of the gadget and
is defined as $\calP_i = \{P_i,Q_i,A_i,B_i,C_i,D_i,E_i\}$ where the
points have constant weights, while $\calC_i$ is the set of initial centers of
the gadget $\calG_i$ and contains exactly two centers. Finally,
$r_i\in \mathbb{R}^+$ and $R_i\in \mathbb{R}^+$ denote respectively the ``inner
radius'' and the ``outer radius'' of the gadget, and their
purpose will be explained later on.
Since the weights of the points do not change between the gadgets, we
will denote the weight of $P_i$ (for any $i>0$) with $w_P$, and similarly
for the other points.

As for the ``leaf'' gadget $\calG_0$, the set $\calP_0$ is composed of only
one point $F$ (of constant weight $w_F$), and $\calC_0$ contains only one
center.

The set of points of the $k$-means istance will be the union of the
(weighted) points from all the gadgets, i.e. $\bigcup_{i=0}^{t-1}
\calP_i$ (with a total of $7(t-1)+1=O(t)$ points of constant weight). 
Similarly, the set of initial centers will be the union of the
centers from all the gadgets, that is $\bigcup_{i=0}^{t-1}\calC_i$
(with a total of $2(t-1)+1=O(t)$ centers).

As we mentioned above, when one of the centers of $\calG_i$ moves to a special
$S^*_{i}$, it will mean that $\calG_i$ fell asleep. For $i>0$ we
define $S^*_{i}$ as the center of mass of the cluster
$\{A_i,B_i,C_i,D_i\}$, while $S^*_0$ coincides with $F$. 

\smallskip

For a gadget $\calG_i$ ($i>0$), we depict the stages (clusterings) it
goes through during any of its day. The entire sequence is shown in
Fig.~2.

\begin{figure}[t]
\begin{center}

\begin{tikzpicture}[scale=1.3]
  \newcommand{\scalefont}[1]{{\small #1}}
  \newcommand{\pointsplot}{%
    \draw[dashed, gray] (0,0) circle (1cm);  
    \draw[dashed, gray] (0,0) circle (1.12cm);

    \draw[dotted, gray] (1, -1.2) -- (1,2);

    \draw[dashed, gray] (0,0) -- (0,-1);
    \draw (0,-0.5) node[right] {\scalefont{$r_i$}};

    \draw[densely dashed, gray] (0,0) -- (-0.5,-1.03);
    \draw (-0.35, -0.35) node {\scalefont{$R_i$}};

    \draw[dotted, gray] (-1.3,0) -- (1.3,0);

    \fill (0,0) circle (0.8pt) node[left] {\scalefont{$P_i$}};
    \fill (0.16,0) circle (0.7pt) node[right] {\scalefont{$Q_i$}};
    \fill (1,-0.7) circle (0.8pt) node[right] {\scalefont{$A_i$}};
    \fill (1,+0.7) circle (0.8pt) node[] {};
    \draw (1.2, +0.6) node {\scalefont{$B_i$}};
    \fill (1,0.85) circle (1.5pt) node[right] {\scalefont{$C_i$}};
    \fill (1,1.7) circle (1.5pt) node[right] {\scalefont{$D_i$}};
    \fill (0,1) circle (2.5pt) node[below left] {\scalefont{$E_i$}};
    
  }
 
  \begin{scope}
    \fill[gray, opacity=0.3, rotate=-30, yshift=1cm] (0,0) ellipse (0.8 and 1.2);
    \fill[gray, opacity=0.3] (1,-0.7) ellipse (0.2 and 0.2);
    \node[diamond,fill, opacity=0.5,scale=0.4] at (0.5, 1) {};
    \pointsplot
    \draw (-1.12, 0) circle (1.5pt) node[below] {\scalefont{$S^*_{i-1}$}};
    \draw (0,-1.2) node[below] {\small \textsc{Morning}};
  \end{scope}
  
  \begin{scope}[xshift=4cm]
    \fill[gray, opacity=0.3, rotate=-20, yshift=1.2cm] (0.25,0) ellipse (0.7 and 1.2);
    \fill[gray, opacity=0.3] (1,-0.7) ellipse (0.2 and 0.2);
    \fill[gray, opacity=0.3] (-0.65,0) ellipse (0.7 and 0.2);
    \node[diamond,fill, opacity=0.5,scale=0.4] at (0.45, 1.1) {};
    \node[diamond,fill, opacity=0.5,scale=0.4] at (-1.1, 0) {};
    \pointsplot
    \fill (-1.12, 0) circle (1.5pt) node[below] {\scalefont{$S^*_{i-1}$}};
    \draw (0,-1.2) node[below] {\small \textsc{1st Call} (pt. I)};
  \end{scope}

  \begin{scope}[xshift=8cm] 
    \fill[gray, opacity=0.3, rotate=+10] (0.8,1.15) ellipse (0.9 and 0.6);
    \node[diamond,fill, opacity=0.5,scale=0.4] at (0.4, 1.2) {};
    \fill[gray, opacity=0.3] (1,0) ellipse (0.3 and 0.75);
    \node[diamond,fill, opacity=0.5,scale=0.4] at (1, 0) {};
    \fill[gray, opacity=0.3] (-0.45,0) ellipse (1.05 and 0.5);
    \node[diamond,fill, opacity=0.5,scale=0.4] at (-1.05, 0) {};
    
    \pointsplot
    \fill (-1.12, 0) circle (1.5pt) node[below] {\scalefont{$S^*_{i-1}$}};
    \draw (0,-1.2) node[below] {\small \textsc{1st Call} (pt. II)};
  \end{scope}

  \begin{scope}[yshift=-4cm, xshift=0cm]
    \fill[gray, opacity=0.3, rotate=+40] (1.3,0.7) ellipse (0.9 and 0.3);
    \node[diamond,fill, opacity=0.5,scale=0.4] at (0.35, 1.22) {};
    \fill[gray, opacity=0.3] (0.9,0) ellipse (1 and 1);
    \node[diamond,fill, opacity=0.5,scale=0.4] at (0.9, 0.8) {};
    
    \pointsplot
    \draw (-1.12, 0) circle (1.5pt) node[below] {\scalefont{$S^*_{i-1}$}};
    \draw (0,-1.2) node[below] {\small \textsc{Afternoon}};
  \end{scope}

  \begin{scope}[yshift=-4cm, xshift=4cm]
    \fill[gray, opacity=0.3, rotate=+40] (1.3,0.7) ellipse (0.9 and 0.3);
    \node[diamond,fill, opacity=0.5,scale=0.4] at (0.35, 1.22) {};
    \fill[gray, opacity=0.3] (1,0) ellipse (0.9 and 1.1);
    \node[diamond,fill, opacity=0.5,scale=0.4] at (0.95, 0.8) {};
    \fill[gray, opacity=0.3] (-0.65,0) ellipse (0.7 and 0.2);
    \node[diamond,fill, opacity=0.5,scale=0.4] at (-1.1, 0) {};

    \pointsplot
    \fill (-1.12, 0) circle (1.5pt) node[below] {\scalefont{$S^*_{i-1}$}};
    \draw (0,-1.2) node[below] {\small \textsc{2nd Call} (pt. I)};
  \end{scope}
  
  \begin{scope}[yshift=-4cm,xshift=8cm]
    \fill[gray, opacity=0.3] (-0.45,0) ellipse (1.05 and 0.5);
    \node[diamond,fill, opacity=0.5,scale=0.4] at (-1.05, 0) {};
    \fill[gray, opacity=0.3] (1,0.5) ellipse (0.3 and 1.5);
    \node[diamond,fill, opacity=0.5,scale=0.4] at (1, 0.9) {};
    \fill[gray, opacity=0.3] (0,1) ellipse (0.3 and 0.3);
    
    \pointsplot
    \fill (-1.12, 0) circle (1.5pt) node[below] {\scalefont{$S^*_{i-1}$}};
    \draw (0,-1.2) node[below] {\small \textsc{2nd Call} (pt. II) / \textsc{Night}};

    \draw[->] (0.45,1.6) -- (0.9,1.1);
    \draw (0.3,1.6) node[] {$S^*_{i}$};
  \end{scope}

\end{tikzpicture}

\parbox{5.5in}{\caption{\small The ``day'' of the gadget $\calG_i$. The diamonds denote the means of the clusters. 
  The locations of the points in figure gives an idea of the actual gadget used in the proof. 
  Also, the bigger the size of a point is, the bigger its weight is. }}
\end{center}
\end{figure}
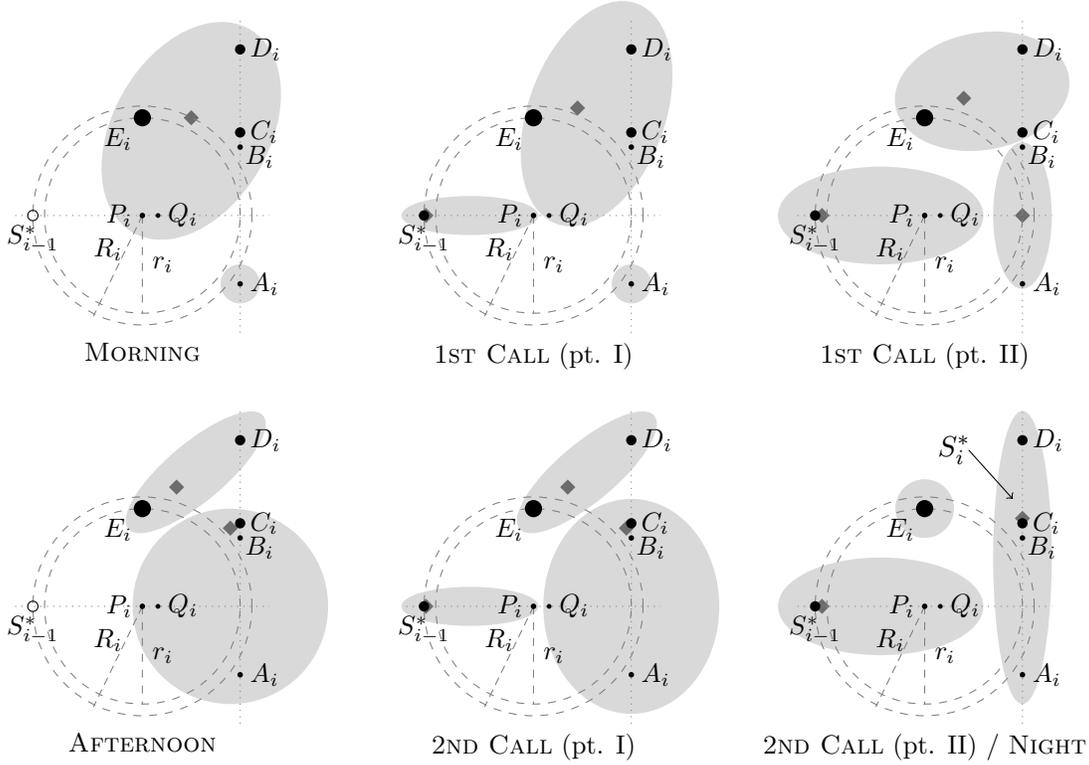

\smallskip

\beginskinny
\noindent\textsc{Morning} This stage takes place right after $\calG_i$
  has been woken up or in the beginning of the entire process. The singleton $\{A_i\}$ is one
  cluster, and the remaining points form the other cluster. In this
  configuration $\calG_i$ is watching $\calG_{i-1}$ and intervenes once it falls asleep.
\endskinny

\beginskinny
\noindent\textsc{1st call} Once $\calG_{i-1}$ falls asleep, $P_i$ will join the
  $\calG_{i-1}$'s cluster with center in $S^*_{i-1}$ (pt. I). At the next step
  (pt. II), $Q_i$ too will join that cluster, and $B_i$ will instead move to the cluster
  $\{A_i\}$. The two points $P_i$ and $Q_i$ are waking up
  $\calG_{i-1}$ by causing a restore of its morning clustering. 
\endskinny

\beginskinny
\noindent\textsc{Afternoon} The points $P_i$, $Q_i$ and $C_i$ will join
  the cluster $\{A_i,B_i\}$. Thus, $\calG_i$ ends up with the clusters
  $\{A_i,B_i,C_i,P_i,Q_i\}$ and $\{D_i,E_i\}$. In this configuration,
  $\calG_i$ is again watching $\calG_{i-1}$ and is ready to wake it up
  once it falls asleep.
\endskinny

\beginskinny
\noindent\textsc{2nd call} Once $\calG_{i-1}$ falls asleep, similarly to the 1st
  call, $P_i$ will join the $\calG_{i-1}$'s cluster with center in
  $S^*_{i-1}$ (pt. I). At the next step (pt. II), $Q_i$ too will join that cluster, and
  $D_i$ will join the cluster $\{A_i,B_i,C_i\}$ (note that the other
  $\calG_{i}$'s cluster is the singleton $\{E_i\}$). Again, $P_i$ and
  $Q_i$ are waking up $\calG_{i-1}$. 
\endskinny

\beginskinny
\noindent\textsc{Night} At this point, the cluster
  $\{A_i,B_i,C_i,D_i\}$ is already formed, which implies that its mean
  is located in $S^*_{i}$: thus, $\calG_{i}$ is sleeping. However, note
  that $P_i$ and $Q_i$ are still in some $\calG_{i-1}$'s cluster and the
  remaining point $E_i$ is in a singleton cluster. In the next step,
  concurrently with the beginning of a possible call from $\calG_{i+1}$
  (see $\calG_{i+1}$'s call, pt.I), the points $P_i$ and $Q_i$ will join the singleton $\{E_i\}$.
\endskinny

The two radiuses of the gadget $\calG_i$ ($i>0$) can be interpreted in the
following way. Whenever $\calG_i$ is watching $\calG_{i-1}$ (either
morning or afternoon), the distance between the point $P$ and its mean will be exactly
$R_i$. On the other hand, the distance between $P_i$ and $S^*_{i-1}$
-- where a $\calG_{i-1}$'s mean will move when
$\calG_{i-1}$ falls asleep -- will be just a bit less than
$R_i$. In this way we guarantee that the waking-up process
will start at the right time. Also, we know that this process will involve $Q_i$ too, and we want
the mean that was originally in $S^*_{i-1}$ to end up at distance more
than $r_i$ from $P_i$. In that step, one of the $\calG_i$'s means will be at
distance exactly $r_i$ from $P_i$, and thus $P_i$ (and $Q_i$ too) will
come back to one of the $\calG_i$'s cluster.

\smallskip

Now we analyze the waking-up process from the point of view of the
sleeping gadget. We suppose that $\calG_i$ ($i>0$) is sleeping and that
$\calG_{i+1}$ wants to wake it up. The sequence is shown in Fig.~3.

\begin{figure}[t]\label{wakeup}
\begin{center}

\begin{tikzpicture}[scale=1.3]
  \newcommand{\scalefont}[1]{{\small #1}}
  \newcommand{\pointsplot}{%
    \draw[dashed, gray] (0,0) circle (1cm);  
    \draw[dashed, gray] (0,0) circle (1.12cm);

    \draw[dotted, gray] (1, -1.2) -- (1,2);

    \draw[dashed, gray] (0,0) -- (0,-1);
    \draw (0,-0.5) node[right] {\scalefont{$r_i$}};

    \draw[densely dashed, gray] (0,0) -- (-0.5,-1.03);
    \draw (-0.35, -0.35) node {\scalefont{$R_i$}};

    \draw[dotted, gray] (-1.3,0) -- (1.3,0);

    \fill (0,0) circle (0.8pt) node[left] {\scalefont{$P_i$}};
    \fill (0.16,0) circle (0.7pt) node[right] {\scalefont{$Q_i$}};
    \fill (1,-0.7) circle (0.8pt) node[right] {\scalefont{$A_i$}};
    \fill (1,+0.7) circle (0.8pt) node[] {};
    \draw (1.2, +0.6) node {\scalefont{$B_i$}};
    \fill (1,0.85) circle (1.5pt) node[right] {\scalefont{$C_i$}};
    \fill (1,1.7) circle (1.5pt) node[right] {\scalefont{$D_i$}};
    \fill (0,1) circle (2.5pt) node[below left] {\scalefont{$E_i$}};
    \fill (3.5,0.9) circle (0.8pt) node[left] {\scalefont{$P_{i+1}$}};
    \fill (3.8,0.9) circle (0.7pt);
    \draw (4,1.1) node {\scalefont{$Q_{i+1}$}};

  }

  \begin{scope}[xshift=0cm]
    \clip (-1.5,-2) rectangle (4.25,2.3);
    \fill[gray, opacity=0.3] (-0.45,0) ellipse (1.05 and 0.5);
    \node[diamond,fill, opacity=0.5,scale=0.4] at (-1.05, 0) {};
    \fill[gray, opacity=0.3] (1,0.5) ellipse (0.3 and 1.5);
    \node[diamond,fill, opacity=0.5,scale=0.4] at (1, 0.9) {};
    \fill[gray, opacity=0.3] (0,1) ellipse (0.3 and 0.3);
    
    \fill[gray, opacity=0.3] (4.5,1.05) ellipse (2 and 0.5);

    \pointsplot
    \fill (-1.2, 0) circle (2pt) node[below] {\scalefont{$S^*_{i-1}$}};

    \draw[->] (0.45,1.6) -- (0.9,1.1);
    \draw (0.3,1.6) node[] {$S^*_{i}$};
    \draw[<->] (1.1,1.9) -- (3.4,1.9);
    \draw (2.3, 2.05) node {\footnotesize $(1-\epsilon)R_{i+1}$};

    \draw (1.4,-1.2) node[below] {\small \textsc{$\calG_i$'s 2nd Call} (pt. II) / \textsc{$\calG_i$'s Night}};

  \end{scope}

  \begin{scope}[xshift=5.2cm]
    \clip (-0.5,-2) rectangle (4.3,2.3);
    
    \fill[gray, opacity=0.3] (0,0.5) ellipse (0.5 and 0.7);
    \node[diamond,fill, opacity=0.5,scale=0.4] at (0, 0.9) {};

    \fill[gray, opacity=0.3, rounded corners] (0.8,1.85)
                     -- (0.8,-1)
                     -- (3.6,0.8)
                     -- (3.6,1.85) -- cycle;
    \node[diamond,fill, opacity=0.5,scale=0.4] at (1.8, 0.9) {};
    
    \fill[gray, opacity=0.3] (4.7,1.05) ellipse (1.05 and 0.7);

    \pointsplot

    \draw (1.4,-1.2) node[below] {\small \textsc{$\calG_{i+1}$'s call} (pt. I)};

  \end{scope}

  \begin{scope}[xshift=-1cm, yshift=-4cm]
    
    \clip (-0.5,-2) rectangle (4.5,2.3);
    
    \fill[gray, opacity=0.3] (0,0.5) ellipse (0.5 and 0.7);
    \node[diamond,fill, opacity=0.5,scale=0.4] at (0, 0.9) {};

    \fill[gray, opacity=0.3, rounded corners] (0.8,1.85)
                     -- (0.8,-1)
                     -- (4.3,0.8)
                     -- (4.3,1.85) -- cycle;
    \node[diamond,fill, opacity=0.5,scale=0.4] at (2.05, 0.9) {};
    \pointsplot

    \draw (1.4,-1.2) node[below] {\small \textsc{$\calG_{i+1}$'s call} (pt. II)};

    \draw[<->] (2.05,1.9) -- (3.4,1.9);
    \draw (2.7, 2.05) node {\footnotesize $(1+\epsilon')r_{i+1}$};
    
  \end{scope}

  \begin{scope}[yshift=-4cm, xshift=5cm]
    
    \clip (-0.5,-2) rectangle (4.5,2.3);
    
    \fill[gray, opacity=0.3, rotate=-30, yshift=1cm] (0,0) ellipse (0.8 and 1.2);
    \fill[gray, opacity=0.3] (1,-0.7) ellipse (0.2 and 0.2);
    \node[diamond,fill, opacity=0.5,scale=0.4] at (0.5, 1.1) {};

    \fill[gray, opacity=0.3] (4.5,1.05) ellipse (2 and 0.5);
    
    \pointsplot

    \draw (2,-1.2) node[below] {\small \textsc{$\calG_{i+1}$'s call} (pt. III) / \textsc{$\calG_i$'s morning}};

  \end{scope}

\end{tikzpicture}

\parbox{5.5in}{\caption{\small \textsc{$\calG_{i+1}$'s call}: how $\calG_{i+1}$ wakes
  up $\calG_i$. The distance between the two gadgets is actually much
  larger than it appears in figure.}}
\end{center}
\end{figure}
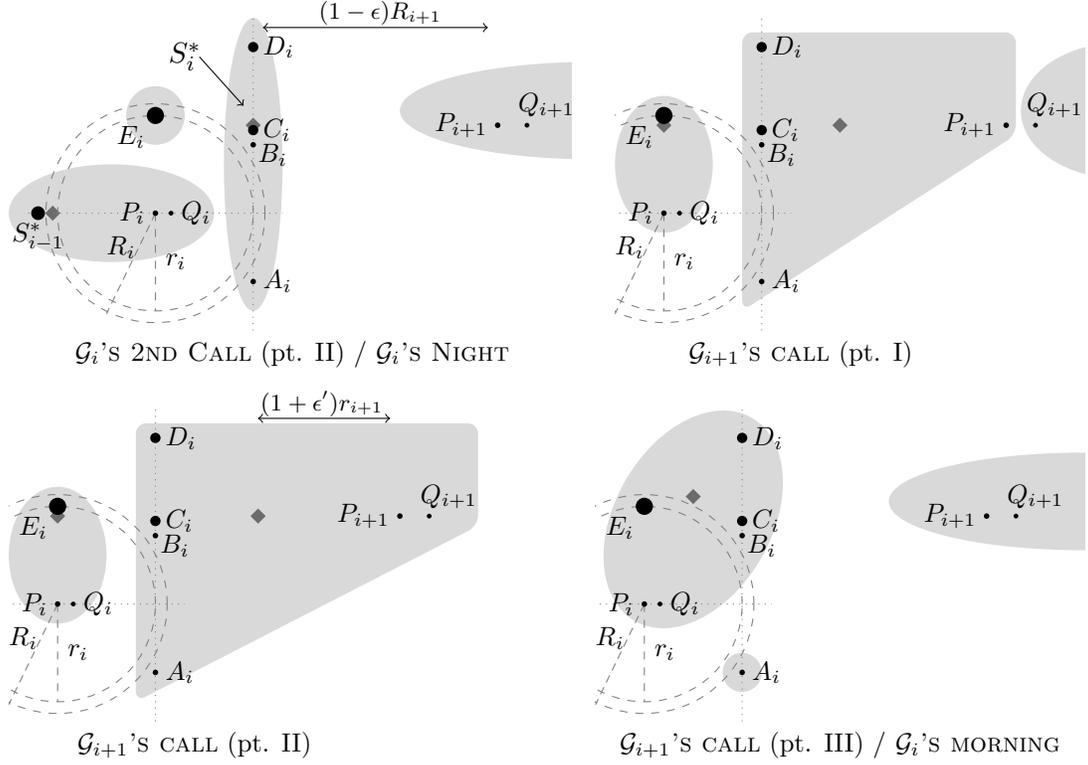

\smallskip

\beginskinny
\textsc{$\calG_{i+1}$'s call} Suppose that $\calG_{i+1}$ started
  to waking up $\calG_{i}$. Then, we know that $P_{i+1}$ joined the
  cluster $\{A_i,B_i,C_i,D_i\}$ (pt. I). However, this does not cause any point
  from this cluster to move to other clusters. On the other hand, as we said
  before, the points $P_i$ and $Q_i$ will ``come back'' to $\calG_{i}$
  by joining the cluster $\{E_i\}$. At the next step (pt. II),
  $Q_{i+1}$ too will join the cluster $\{A_i,B_i,C_i,D_i,P_{i+1}\}$. 
  The new center will be in a position such that, in one more step
  (pt. III), $B_i,C_i$ and $D_i$ will move to the cluster
  $\{P_i,Q_i,E_i\}$. 
  Also we know that at that very same step, $P_{i+1}$ and $Q_{i+1}$ will come back to some
  $\calG_{i+1}$'s cluster: this implies that $\calG_{i}$ will end up
  with the clusters $\{B_i,C_i,D_i,E_i,P_i,Q_i\}$ and $\{A_i\}$, which
  is exactly the morning clustering: $\calG_i$ has been woken up.
\endskinny

\noindent As for the ``leaf'' gadget $\calG_0$, we said that it will fall asleep right after
it has been woken up by $\calG_1$. Thus we can describe its day in the
following way:

\smallskip

\beginskinny
\textsc{Night} There is only one cluster which is the singleton
  $\{F\}$. The center is obviously $F$ which coincides with
  $S^*_0$. In this configuration $\calG_0$ is sleeping. 

\smallskip

\textsc{$\calG_1$'s call} The point $P_{1}$ from $\calG_1$ joins
the cluster $\{P_{0}\}$ and in the next step $Q_1$ will join the same
cluster too. After one more step, both $P_1$ and $Q_1$ will come back
to some $\calG_1$'s cluster, which implies that the $\calG_0$'s cluster is the
singleton $\{F\}$ again. Thus $\calG_0$, after having been
temporarily woken up, fell asleep again. 
\endskinny

\subsection{Formal Construction}
We start giving the distances between the points in a single gadget (intra-gadget). 
Afterwards, we will give the distances between two consecutive gadgets (inter-gadget).
Henceforth $x_{A_i}$ and $y_{A_i}$ will denote respectively the $x$-coordinate and
$y$-coordinate of the point $A_i$, and analogous notation will be used
for the other points. Also, for a set of points $\calS$, we define its
total weight $w_{\calS}=\sum_{x\in\calS}w_x$, and its mean will be
denoted by $\mu(\calS)$, i.e. $\mu(\calS) = \frac{\sum_{x\in\calS} w_x\cdot x}{w_{\calS}}$. 
We suppose that all the weights $w_P,w_Q,w_A,\ldots$ have been fixed
to some positive integer values, and that $w_A=w_B$ and $w_F=w_A+w_B+w_C+w_D$.

\smallskip

We start describing the distances between points for a non-leaf gadget. For
simplicity, we start defining the location of the points for an
hypotetical ``unit'' gadget $\hat{\calG}$ that has unitary inner
radius (i.e. $\hat{r} = 1$) and is centered in the origin
(i.e. $\hat{P}=(0,0)$). Then we will see how to define a gadget
$\calG_i$ (for any $i>0$) in terms of the unit gadget $\hat{\calG}$.

The outer radius is defined as $\hat{R} = (1+\delta)$ and also we let the point $\hat{Q}$ be
$\hat{Q}=(\lambda,0)$. The values $0<\delta<1$  and $0<\lambda<1$ are
constants whose value will be assigned later. 
The point $\hat{E}$ is defined as $\hat{E}=(0,1)$.

The remaining points are aligned on the vertical line with
$x$-coordinate equals to $1$ (formally, $x_{\hat{A}}=x_{\hat{B}}=x_{\hat{C}}=x_{\hat{D}}=1$).
As for the $y$-coordinates, we set $y_{\hat{A}}=-1/2$ and $y_{\hat{B}}=1/2$. 

The value $y_{\hat{C}}$ is uniquely defined by imposing 
$y_{\hat{C}}>0$ and that the mean of the cluster
$\calM=\{\hat{A},\hat{B},\hat{C},\hat{P},\hat{Q}\}$ is at distance $\hat{R}$ from $\hat{P}$. 
Thus, we want the positive $y_{\hat{C}}$ that satisfies the equation
$||\mu(\calM)||=\hat{R}$, which can be rewritten as
$$
\left(\frac{w_A+w_B+w_C+w_Q\lambda}{w_\calM}\right)^2 + \left(\frac{w_C
  y_{\hat{C}}}{w_\calM}\right)^2  = (1+\delta)^2
$$
where we used the fact that $w_A y_{\hat{A}} + w_B y_{\hat{B}} = 0$ when
$w_A=w_B$.

We easily obtain the solution
$$
y_{\hat{C}} = \frac{1}{w_C} \sqrt{(w_\calM (1+\delta))^2 - (w_A+w_B+w_C+w_Q\lambda)^2}
$$
Note that the value under the square root is always positive because $\lambda<1$.

It remains to set $y_{\hat{D}}$. Its value is uniquely defined by imposing
$y_{\hat{D}}>0$ and that the mean of the cluster
$\calN=\{\hat{B},\hat{C},\hat{D},\hat{E},\hat{P},\hat{Q}\}$ is at distance $\hat{R}$ from
$\hat{P}$. Analogously to the previous case, $y_{\hat{D}}$ is the positive value
satisfying $||\mu(\calN)||=\hat{R}$, which is equivalent to
$$
\left(\frac{w_B+w_C+w_D+w_Q\lambda}{w_\calN}\right)^2 + \left(\frac{w_D
  y_{\hat{D}} + w_B (1/2) + w_C y_{\hat{C}} + w_E }{w_\calN}\right)^2  = (1+\delta)^2
$$
Now, since the equation $a^2+(b+x)^2=c^2$ has the solutions
$x=\pm \sqrt{c^2-a^2}-b$, we obtain the solution
$$
y_{D_i} = \frac{1}{w_D}\left|\sqrt{(w_\calN (1+\delta))^2-(w_B+w_C+w_D+w_Q\lambda)^2} - w_B/2 -w_C y_{\hat{C}} -w_E\right|
$$
Again, the term under the square root is always positive.

Finally, we define $\hat{S}^*$ in the natural way as $\hat{S}^* = \mu\{\hat{A},\hat{B},\hat{C},\hat{D}\}$.

\smallskip

Now consider a gadget $\calG_i$ with $i>0$. Suppose to have fixed the
inner radius $r_i$ and the center $P_i$. Then we have the outer radius
$R_i=(1+\delta)r_i$, and we define the location of the
points in terms of the unit gadget by scaling of $r_i$ and translating by $P_i$ in following way:
$A_i = P_i + r_i\hat{A}$, $B_i = P_i + r_i\hat{B}$, and so on for the other points.  

As for the gadget $\calG_0$, there are no intra-gadget distances to be defined,
since it has only one point $F$.

\smallskip

For any $i\ge 0$, the intra-gadget distances in $\calG_i$ have been
defined (as a function of $P_i$, $r_i$, $\delta$ and $\lambda$). Now 
we define the (inter-gadget) distances between the points of two consecutive gadgets
$\calG_i$ and $\calG_{i+1}$, for any $i\ge 0$. We do this by giving
expliciting recursive expressions for $r_i$ and $P_i$.

For a point $\hat{Z}\in \{\hat{A},\hat{B},\hat{C},\hat{D}\}$, we
define the ``stretch'' of $\hat{Z}$ (from $\hat{S^*}$ with respect
to $\mu\{\hat{E},\hat{P},\hat{Q}\}$) as 
$$
\sigma(\hat{Z}) = \sqrt{d^2(\hat{Z},\mu\{\hat{E},\hat{P},\hat{Q}\})-d^2(\hat{Z},\hat{S}^*)}
$$
The stretch will be a real number (for all points
$\hat{A},\hat{B},\hat{C},\hat{D}$), 
given the values $\lambda$, $\delta$ and the weights used in the construction.

We set the inner radius $r_0$ of the leaf gadget $\calG_0$ to a positive arbitrary value, and for any $i\ge 0$, we define
\begin{eqnarray}
\label{radius}
r_{i+1} = \frac{r_i}{1+\delta}\frac{w_F+w_P+w_Q}{w_P+(1+\lambda)w_Q}\sigma(\hat{A})
\end{eqnarray}
where we remind that $w_F = w_A+w_B+w_C+w_D$.

Now recall that $S^*_i=\mu\{A_i,B_i,C_i,D_i\}$ for any $i>0$, and $S^*_0=\mu\{F\}=F$.
Assuming to have fixed the point $F$ somewhere in the plane, we define for any $i>0$ 
\begin{eqnarray}
\label{interdistance} x_{P_{i}} & = & x_{S^*_{i-1}} + R_{i}(1-\epsilon)\\
\nonumber y_{P_{i}} & = & y_{S^*_{i-1}}
\end{eqnarray}
where $0<\epsilon<1$ is some constant to define. Note that now the instance is completely defined in function of $\lambda$,
$\delta$, $\epsilon$ and the weights. We are now ready to prove the lower bound.

\subsection{Proof}
We assume that the initial centers -- that we seed $k$-means with --
correspond to the means of the ``morning clusters'' of each gadget
$\calG_i$ with $i>0$. Namely, the initial centers are $\mu\{A_i\}$,
$\mu\{B_i,C_i,D_i,E_i,P_i,Q_i\}$ for all $i>0$, in addition to the
center $\mu\{F\}=F$ for the leaf gadget $\calG_0$. 

\smallskip 

In order to establish our result, it is enough to show that there
exist positive integer values $w_A,w_B,w_C,w_D,w_E,w_F,w_P,w_Q$ (with $w_A=w_B$)
and values for $\lambda$, $\delta$ and $\epsilon$, such that the behavior of
$k$-means on the instance reflects exactly the clustering transitions described in
Section~\ref{defs}.
The chosen values (as well as other derived values used later in the analysis)
are in Table~\ref{values}.
The use of rational weights is not restrictive, because the mean of a cluster
(as well as $k$-means' behavior) does not change if we multiply the
weights of its points by the same factor -- in our case it is enough
to multiply all the weights by $100$ to obtain integer weights.

Finally, for the value of $\epsilon$, we impose
$$
0 < \epsilon < \min\left\{\frac{d^2(\hat{S}^*,\hat{C})}{(1+\delta)^2}, 
                          \frac{\lambda}{1+\delta}, 
                          \frac{\sigma(\hat{A})-\sigma(\hat{B})}{\sigma(\hat{A})},
                          1-\frac{(1+\lambda w_Q)(w_F+w_P+w_Q)}{(1+\delta)w_F}
                          \right\}
$$

Throughout the proof, we will say that a point $Z$ in a cluster
$\calC$ is \emph{stable} with respect to (w.r.t) another cluster
$\calC'$, if $d(Z,\mu(\calC))<d(Z,\mu(\calC'))$. Similarly, a point $Z$ in
a cluster $\calC$ is stable if $Z$ is stable w.r.t. any $\calC'\neq
\calC$. Also, similar definitions of stability extends to a cluster
(resp. clustering) if the stability holds for all the points
in the cluster (resp. for all the clusters in the clustering).

We consider an arbitrary gadget $\calG_i$ with $i>0$ in any stage of
its day (some clustering), and we show that the steps that $k$-means
goes through are exactly the ones described in Section~\ref{defs} for
that stage of the day (for the chosen values of $\lambda,\delta,\epsilon$ and weights).
For the sake of convenience and w.l.o.g, we assume that $\calG_i$ has
unitary inner radius (i.e. $r_i=\hat{r}=1$ and
$R_i=\hat{R}=(1+\delta)$) and that $P_i$ is in the origin (i.e. $P_i=(0,0)$).

\begin{table}[t]\footnotesize
  \begin{center}
  \begin{tabular}{|l|l|l|l|}
  \hline
  Chosen values       & Unit gadget                          & Other derived values used in the proof\\
  \hline
  $\delta = 0.25$     & $\hat{r} = 1$                        & $(0.1432, 1.0149)\preceq N \preceq (1.44,1.015)$\\
  $\lambda = 10^{-5}$ & $\hat{R} =(1+\delta) = 1.025$        & $(0.9495, 0.386)\preceq M \preceq (0.9496,0.3861)$\\
  $w_P = 1$           & $\hat{P} = (0,0)$                    & $1.003 \le \alpha \le 1.004$\\
  $w_Q = 10^{-2}$     & $\hat{Q} = (\lambda,0) = (10^{-5},0)$& $1.0526\le \beta \le 1.05261$\\
  $w_A = 4$           & $\hat{A} = (1, -0.5)$                & $0.99\le \gamma \le 0.99047$\\               
  $w_B = 4$           & $\hat{B} = (1, 0.5)$                 & $1.0003\le \sigma(\hat{A}) \le 1.0004$   \\
  $w_C = 11$          & $(1,0.70223)\preceq\hat{C}\preceq (1,0.70224)$  & $1.0001\le \sigma(\hat{B}) \le 1.0002$   \\
  $w_D = 31$          & $(1,1.35739)\preceq\hat{D}\preceq (1,1.3574)$ & $1\le \sigma(\hat{C}) \le 1.0001$   \\
  $w_E = 274$         & $\hat{E} = (0,1)$                    & $0.9999\le \sigma(\hat{D}) \le 0.99992$   \\
  \hline
  \end{tabular}
  \caption{The relation $\preceq$ denotes the less-or-equal component-wise relation.}
  \label{values}
\end{center}
\end{table}

\vspace{-0.1cm}
\[\noindent\textsc{Morning}\] 
We need to prove that the morning clustering of
$\calG_i$ is stable assuming that $\calG_{i-1}$ is not sleeping. Note that this assumption
implies that $i>1$ since the gadget $\calG_0$ is always sleeping when
$\calG_1$ is in the morning. 
Since the singleton cluster $\{A_i\}$ is trivially stable, we just need to
show that $\calN = \{B_i,C_i,D_i,E_i,P_i,Q_i\}$ is stable. It is easy
to understand that it suffices to show that $B_i$, $Q_i$ and $P_i$ are
stable w.r.t $\{A_i\}$ (the other points in $\calN$ are further from $A_i$),
and that $P_i$ is stable w.r.t any $\calG_{i-1}$'s cluster.
Letting $N=\mu(\calN)$, we have $x_N =(w_B+w_C+w_D+\lambda w_Q)/w_{\calN}$,
and $y_N =\sqrt{(1+\delta)^2-x_N^2}$. 

The point $P_i$ is stable w.r.t. $\{A_i\}$, since $d(P_i,N) = (1+\delta) < \sqrt{1^2+(0.5)^2} =
d(P_i,A_i)$. To prove the same for $Q_i$, note that $d(Q_i,A_i) =
\sqrt{(1-\lambda)^2+(0.5)^2} > \hat{R}$, while on the other hand
$x_N>x_{Q_i}$ implies $d(Q_i,N)<\hat{R}$.

As for $B_i$, $d^2(B_i,N) = (x_B-x_N)^2+(y_B-y_N)^2 =
||B_i||^2+\hat{R^2}-2(x_Nx_{B_i}+y_Ny_{B_i})$. Thus, the
inequality $d(B_i,N)<d(B_i,A_i)=1$ simplifies to
$5/4+\hat{R}^2-2x_N-y_N<1$, which can be checked to be valid.

It remains to prove that $P_i$ is stable w.r.t. any $\calG_{i-1}$'s
cluster. It is easy to understand that, in any stage of $\calG_{i-1}$'s day (different
from the night), the distance from any $\calG_{i-1}$'s center to $P_i$
is more than the distance between $C_{i-1}$ and $P_i$. We observe that
$d^2(P_i,C_{i-1}) = (x_{P_i}-x_{S^*_{i-1}})^2 +
d^2(S^*_{i-1},C_{i-1}) = R^2_i(1-\epsilon)^2 +
\hat{r}d^2(\hat{S}^*_{i-1},\hat{C})$, using (\ref{interdistance}).
The assumption $\epsilon<d^2(\hat{S}^*,\hat{C})/(1+\delta)^2$ directly
implies $d^2(P_i,C_{i-1})>(1+\delta)=d(P_i,N)$.

\[\noindent\textsc{1st Call}\]
We start analyzing the part I of this stage. Since we are assuming
that $\calG_{i-1}$ is sleeping, there must be some $\calG_{i-1}$'s cluster $\calC$ with center in
$S^*_{i-1}$ (note that $\calG_{i-1}$ can be the leaf gadget $\calG_0$
as well). 
By (\ref{interdistance}) we have $d(P_i,S^*_{i-1})< R_i$,
and so $P_i$ will join $\calC$. We claim that $Q_i$ (any other
$\calG_i$'s point is implied) is instead stable, i.e. $d(Q_i,N)<
d(Q_i,S^*_{i-1})$. 
We already know that $d(Q_i,N)< \hat{R}$, so we show
$d(Q_i,S^*_{i-1})>\hat{R}$. 
Using (\ref{interdistance}), we have $\hat{R}(1-\epsilon)+\lambda \hat{r}>\hat{R}$,
which holds for $\epsilon < \lambda/(1+\delta)$.

\smallskip

We now analyze the next iteration, i.e. the part II of this stage.
We claim that $Q_i$ will join
$\calC\cup\{P_i\}$, and $B_i$ will join $\{A_i\}$. 
To establish the former, we show that $d(Q_i,\mu(\calN'))> \hat{R}$ where $\calN'=\calN-\{P_i\}$.
Since $P_i$ is in the origin, we can write $N'=\alpha N$ with
$\alpha=w_{\calN}/w_{\calN'}$.
Thus, the inequality we are interested in is $(\lambda-\alpha x_{N})^2+(\alpha
y_{N})^2>\hat{R}^2$ which can rewritten as $(\alpha^2-1)\hat{R}>2\lambda\alpha x_N$.
Finally, since $\alpha>1,\hat{R}>1$ and $x_N<1$, the inequality is
implied by $\alpha(1-2\lambda)>1$, which holds for the chosen values.
 
It remains to prove that $B_i$ is not stable w.r.t. $\{A_i\}$,
i.e. $d(B_i,N')>d(B_i,A_i)=1$. Again, starting with the inequality 
$(1-\alpha x_{N})^2+(1/2-\alpha y_{N})^2>1$, we get the equivalent inequality 
$1/4+\alpha^2\hat{R}>\alpha(2x_N+y_N)$, which is easy to verify.

Finally, we prove that $C_i$ is instead stable w.r.t. $N'$. Similarly we get 
$x_{C_i}^2+y_{C_i}^2+\alpha^2\hat{R}^2-2\alpha(x_N x_{C_i}+y_N y_{C_i})< (y_{A_i}-y_{C_i}^2)$, 
which is implied by $3/4+\alpha^2\hat{R}^2<y_{C_i}(1+2\alpha y_N)$.

\[\noindent\textsc{Afternoon}\]
The last stage ended up with the $\calG_i$'s clusters
$\calN''=\{C_i,D_i,E_i\}$ and $\{A_i,B_i\}$, since $P_i$ and $Q_i$
both joined the cluster $\calC$ of $\calG_{i-1}$. We claim that, at this point,
$P_i,Q_i$ and $C_i$ are not stable and will all join the cluster
$\{A_i,B_i\}$. 

Let $\calC'=\calC\cup\{P_i,Q_i\}$; note that the total weight $w_{\calC'}$
of the cluster $\calC'$ is the same if $\calG_{i-1}$ is the leaf gadget $\calG_0$ or not,
since by definition of $w_{\calC}=w_F=w_A+w_B+w_C+w_D$. We start showing that
$d(P_i,\mu(\calC'))>\hat{r}=1$ which proves that the claim is true for
$P_i$ and $Q_i$. By defining $d=x_{P_i}-x_{S^*_{i-1}}$, the inequality can be rewritten as
$d - (w_Pd + w_Q(d+\lambda))/w_{\calC'} > 1$, which by (\ref{interdistance})
is equivalent to $(1-\epsilon)(1+\delta)w_{\calC}/w_{\calC'}> 1+\lambda w_Q$.
It can be checked that $(1+\delta)w_{\calC}/w_{\calC'}>1+\lambda w_Q$ and 
the assumption on $\epsilon$ completes the proof.

Now we prove that $C_i$ is not stable w.r.t to $\{A_i,B_i\}$, by
showing that $d(C_i,N'')>y_{C_i}$ where $N''=\mu(\calN'')$. Note that
the inequality is implied by $x_{C_i}-x_{N''}>y_{C_i}$, which is
equivalent to $w_E/w_{\calN''} > y_{C_i}$ that holds for the chosen values.

\smallskip 

At this point, analogolously to the morning stage, we want to show that
this new clustering is stable, assuming that $\calG_{i-1}$ is not
sleeping. Note that the analysis in the morning stage directly implies that
$P_i$ is stable w.r.t any $\calG_{i-1}$'s cluster. It can be shown as
well that $P_i$ is stable w.r.t to $\calN''' = \{D_i,E_i\}$, and $D_i$ is stable
w.r.t. $\calM=\{A_i,B_i,C_i,P_i,Q_i\}$ (other points' stability is implied).

\[\noindent\textsc{2nd Call}\]
For the part I of this stage, i.e. we assume $\calG_{i-1}$ is
sleeping, and so there is some $\calG_{i-1}$'s cluster $\calC$ with center in
$S^*_{i-1}$. Similarly to the 1st call (part I), $P_i$ will join $\calC$. 
The point $Q_i$ is instead stable, since we proved
$d(Q_i,S^*_{i-1})>\hat{R}$, while $x_M>x_{Q_i}$ implies $d(Q_i,M)<\hat{R}$.

\smallskip

We now analyze the next iteration, i.e. the part II of this stage.
We claim that $Q_i$ will join $\calC\cup\{P_i\}$, and $D_i$ will join
$\calM' = \calM-\{P_i\}$. This can be proven analogously to the
part II of the first call, by using $M'=\mu(\calM')=\beta M$, where 
$\beta=w_{\calM}/w_{\calM'}$.

\[\noindent\textsc{Night}\]
The last stage leaves us with the clusters $\{A_i,B_i,C_i,D_i\}$ and
the singleton $\{E_i\}$. We want to prove that in one iteration $P_i$
and $Q_i$ will join $\{E_i\}$. In the afternoon stage, we already proved
that $d(P_i,\mu(\calC'))>\hat{r}$, and since $d(P_i,A_i)=\hat{r}=1$, the
point $P_i$ will join $\{E_i\}$. For the point $Q_i$, we have
$d(Q_i,\mu(\calC'))=d(P_i,\mu(\calC'))+\lambda>\hat{r}+\lambda$,
while $d(Q_i,E_i)=\sqrt{\hat{r}^2+\lambda^2}<\hat{r}+\lambda$. 
Thus, the point $Q_i$, as well as $P_i$, will join $\{E_i\}$.

\[\noindent\textsc{$\calG_{i+1}$'s call}\]
In this stage, we are analyzing the waking-up process from the point
of view of the sleeping gadget. We suppose that $\calG_i$ ($i>0$) is sleeping and that
$\calG_{i+1}$ wants to wake it up. 

We start considering the part I of this stage, when only $P_{i+1}$
joined the cluster $\calS=\{A_i,B_i,C_i,D_i\}$. Let
$\calS'=\calS\cup\{P_{i+1}\}$.
We want to verify that the points in $\calS$
are stable w.r.t. $\{E_i,P_i,Q_i\}$, i.e. that for each
$\hat{Z}\in\calS$, $d(\hat{Z},\mu(\calS')) < d(\hat{Z},\mu\{E_i,P_i,Q_i\})$. 
This inequality is equivalent to $d(\hat{S}^*,\mu(\calS')) <
\sigma(\hat{Z})$, and given the ordering of the stretches, it is
enough to show it for $\hat{Z}=\hat{D}$. 
By (\ref{interdistance}), we have that $d(\hat{S}^*,\mu(\calS')) =
(1-\epsilon)R_{i+1}w_P/w_{\calS'}$, and using (\ref{radius}) we get 
$d(\hat{S}^*,\mu(\calS')) = \hat{r}(1-\epsilon)\gamma\sigma(\hat{A})$ where
$\gamma=(w_P/w_{\calS'})(w_{\calS'}+w_Q)/(w_P+(1+\lambda)w_Q)$. Finally, it is easy to verify
that $\gamma\sigma(\hat{A})<\sigma(\hat{D})$.

\smallskip

In the part II of this stage, $Q_{i+1}$ joined $\calS'$. Let $\calS''=\calS'\cup\{Q_{i+1}\}$.. 
We want to verify that all the points in $\calS$ but $A$ will
move to the cluster $\{E_i,P_i,Q_i\}$. 

We start showing that
$d(A_i,\mu(\calS''))<d(\hat{Z},\mu\{E_i,P_i,Q_i\})$. 
This inequality is equivalent to $d(\hat{S}^*,\mu(\calS'')) <
\sigma(\hat{A})$, and we have
$d(\hat{S}^*,\mu(\calS''))=(1-\epsilon)R_{i+1}(w_P+(1+\lambda)w_Q)/(w_P+w_Q+w_F)$.
Using (\ref{radius}) to substitute $R_{i+1}$, we get $d(\hat{S}^*,\mu(\calS'')) =
(1-\epsilon)\sigma(\hat{A})$, which proves that $A_i$ will not change
cluster.

Similarly, we want to prove that, for $\hat{Z}\in\calS$,
$\hat{Z}\neq\hat{A}$, it holds that
$d(\hat{S}^*,\mu(\calS'')) = (1-\epsilon)\sigma(\hat{A}) > \sigma(\hat{Z})$. Given the ordering of
the stretches, it suffices to show it for $\hat{Z}=\hat{B}$. Recalling that
$\epsilon<(\sigma(\hat{A})-\sigma(\hat{B}))/\sigma(\hat{A})$, the
proof is concluded.

\subsection{Extensions}
The proof in the previous section assumed that the set of initial
centers correspond to the means of the ``morning clusters'' for each gadget
$\calG_i$ with $i>0$. A common initialization for $k$-means is to choose the 
set of centers among the data points. We now briefly explain how to modify our 
instance so to have this property and the same number of iterations. 

Consider the unit gadget $\hat{\calG}$ for simplicity. One of the
center will be the point $\hat{E}$. In the beginning we want all the points of
$\hat{\calG}$ except $\hat{A}$ to be assigned to $\hat{E}$. To obtain
this, we will consider two new data points each with a center on it.
Add a point (and center) $\hat{I}$ with $x_{\hat{I}}=x_{\hat{A}}=1$ and such that
$y_{\hat{A}}-y_{\hat{I}}$ is slightly less than $d(\hat{A},\hat{E})$. In
this way $\hat{A}$ will be assigned to this center. 
Also, we add another point (and center) $\hat{J}$ very close to
$\hat{I}$ (but further from $\hat{A}$) so that, when $\hat{B}$ joins the cluster $\{\hat{I}\}$
moving the center towards itself, the point $\hat{I}$ will move to the
cluster $\{\hat{J}\}$. By modifying in this way all the gadgets in the
instance, we will reach the morning clustering of each gadget in two
steps. 
Also it is easy to check that the new points do not affect the following steps. 

\smallskip

Har-Peled and Sadri \cite{harpeled} conjectured that, for any
dimension $d$, the number of iterations of $k$-means might be 
bounded by some poynomial in the number of point $n$ and the spread
$\Delta$ ($\Delta$ is ratio between the largest and the smallest
pairwise distance).

This conjecture was already disproven in \cite{arthur} for
$d=\Omega(\sqrt{n})$. By using the same argument, we
can modify our construction to an instance in $d=3$ dimension having
linear spread, for which $k$-means requires $2^{\Omega(n)}$ iterations. Thus, the
conjecture does not hold for any $d\ge 3$.

\section{Conclusions and further discussion}
We presented how to construct a $2$-dimensional instance with $k$ clusters
for which the $k$-means algorithm requires $2^{\Omega(k)}$
iterations. For $k=\Theta(n)$, we obtain the lower bound $2^{\Omega(n)}$.
Our result improves the best known lower bound \cite{arthur} in terms of number of iterations
(which was $2^{\Omega(\sqrt{n})}$), as well as in terms of
dimensionality (it held for $d=\Omega(\sqrt{n})$).

We observe that in our construction each gadget uses a constant number
of points and wakes up the next gadget twice. For $k=o(n)$, we could
use $\Theta(n/k)$ points for each gadget, and it would be interesting
to see if one can construct a gadget with such many points that is able
to wake up the next one $\Omega(n/k)$ times. Note that this would give
the lower bound $(n/k)^{\Omega(n/k)}$, which for $k=n^c$ ($0<c<1$),
simplifies to $n^{\Omega(k)}$. This matches the optimal upper bound
$O(n^{kd})$, as long as the construction lies in a constant number of dimensions.

A polynomial upper bound for the case $d=1$ has been recently proven
in the smoothed regime \cite{manthey}. It is natural to ask if this
result can be extended to the ordinary case.

\section*{Acknowledgements}
We greatly thank Flavio Chierichetti and Sanjoy Dasgupta
for their helpful comments and discussions.  We also thank David
Arthur for having confirmed some of our intuitions on the proof in \cite{arthur}.

\end{document}